\documentclass[12pt, letterpaper]{article}
\usepackage[T1]{fontenc}    
\usepackage{hyperref}       
\usepackage{url}            
\usepackage{booktabs}       
\usepackage{amsfonts}       
\usepackage{nicefrac}       
\usepackage{microtype}      
\usepackage{lipsum}		
\usepackage{graphicx}

\usepackage{doi}

\usepackage{graphicx}        
\usepackage[bottom]{footmisc}

\usepackage{caption}
\usepackage{subcaption}
\usepackage[linesnumbered,ruled,vlined]{algorithm2e}
\usepackage{amsmath}
\usepackage{amsfonts}

\usepackage[utf8]{inputenc}

\author{
  Kevser Şahinbaş\\
  \texttt{ksahinbas@medipol.edu.tr}\\
  Istanbul Medipol University,  Istanbul, Turkey
  \and
  Ferhat Ozgur Catak\\
  \texttt{ozgur@simula.no}\\
  Simula Research Laboratory, Fornebu, Norway
}

\usepackage[square,sort,comma,numbers]{natbib}

\usepackage{graphicx}

\begin{document}
\title{Secure Multi-Party Computation based Privacy Preserving Data Analysis in Healthcare IoT Systems}

\maketitle

\abstract{Recently, many innovations have been experienced in healthcare by rapidly growing Internet-of-Things (IoT) technology that provides significant developments and facilities in the health sector and improves daily human life. The IoT bridges people, information technology and speed up shopping. For these reasons, IoT technology has started to be used on a large scale. Thanks to the use of IoT technology in health services, chronic disease monitoring, health monitoring, rapid intervention, early diagnosis and treatment, etc. facilitates the delivery of health services. However, the data transferred to the digital environment pose a threat of privacy leakage. Unauthorized persons have used them, and there have been malicious attacks on the health and privacy of individuals. In this study, it is aimed to propose a model to handle the privacy problems based on federated learning. Besides, we apply secure multi party computation. Our proposed model presents an extensive privacy and data analysis and achieve high performance.}

\section{Introduction}
Cloud Computing technology, one of the rapidly developing technologies in the information technologies, provides secure and location-independent storage, computing and various application services. Besides, with the production of new generation hardware with different sensors, low cost and efficient energy consumption, the Internet of Things that transform any sensitive data into network data by wireless sensor networks, has been widely used in the detection, monitoring, control and smart management system \cite{lin2017survey,guo2019deep,10.7717/peerj-cs.346}. New generation applications that benefit from these two technologies provide important services in different fields. For example, the advantage of the IoT and cloud computing technologies have been extensively adopted in healthcare, such as remote patient care and hand hygiene monitoring systems \cite{yu2012smart, karimpour2019iot,SAHINBAS2021451}. Health devices; Information about the devices such as their instant location, calibration status and measures against theft can be followed instantly. In addition, significant progress has been made in monitoring the health of people outside of the hospital. People in their homes, workplaces or anywhere can communicate with health service providers and organizations with IoT technology and access healthcare services remotely. With the development of 5G technology, it is thought that IoT will take a more active role in our life. IoT technology is becoming the most powerful and flexible technology on the market. Therefore, the market is currently flooded with various devices and systems in multiple fields of technology development and commercialization. With the increasing number of such products and services, consumers are starting to perceive the value of privacy and security of their personal data in new ways \cite{electronics9020229}.  In addition to privacy concerns, consumers are concerned about the growing usage of internet-based devices and the resulting surveillance in public regarding personal data. There is increasing acceptance of data protection as a fundamental right in developed countries, and there is growing public awareness regarding personal data privacy. Therefore, the market is highly concentrated in the area of data protection and privacy with IoT data.

\begin{figure}[htbp!]
    \centering
    \includegraphics[width=1.0\linewidth]{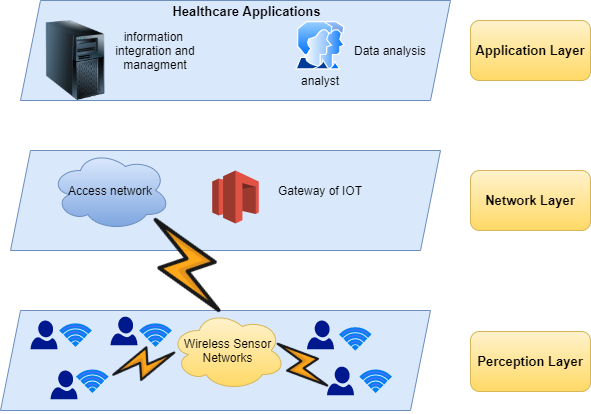}
    \caption{The architecture of healthcare IoT systems}
    \label{fig:iot}
\end{figure}

Real-time and accurate data collection from sensors is one of the critical functions in the Internet of Things. Generally, in sensor-based systems, an asset's environment, structure, or various properties are detected by sensors and sent to a base station and from there to a data processing center over WiFi, a wired communication infrastructure, or multi-hop connections. After the data reaches the processing center, intelligent decisions can be made based on the perceived information or the quality of the services provided. Figure \ref{fig:iot} indicates the general system architecture of IoT systems in healthcare \cite{vijayalakshmi2018secured}.

The number of devices connected to the internet has been increasing recently, and these devices are in homes and workplaces with the security risks they contain. Connecting billions of devices, IoT encompasses various things from wearables, smart home appliances, vehicle sensors and medical devices. With the ubiquity of IoT technology, it is also exposed to many cyberattacks. Those who take control of the devices can dangerously use these devices. This situation requires the use of security systems in IoT technology. Thus, privacy and security issues must be considered and addressed. To meet such requirements, we will propose a model that overcomes privacy problems by applying federated learning (FL). 

We summarized some of the surveyed researches in the literature in relation to the privacy-preserving healthcare IoT data. Gou et al. \cite{guo2019deep} presents a mutual privacy-preserving k-means strategy (M-PPKS) model that applies homomorphic encryption to handle privacy issues. Sage \cite{lin2009sage} provides a privacy-preserving outline by implementing privacy-sensitive data transmission. It ensures content confidentiality by using encryption methods in its work. It provides the anonymity of the receiver using the broadcast strategy that allows hiding the potential doctor. Huang et al.\cite{huang2012privacy} provide a privacy plan for e-health services by tracking the patients' physical condition to obligate anonymous authentication. By encrypting data exchanged using Elliptic Curve Cryptography (ECC), their plans also provide content-driven privacy. The study ensures unlinkability between the identity and the biometric information of each patient. \cite{lu2012spoc} provides an Opportunistic Computing framework that maintains Secure Privacy for the m-Healthcare emergency. The purpose of the study is to aid patients monitor and protect personal health data in pervasive environments. It provides content privacy based on cryptographic primitives. Zhang et al. \cite{zhang2016light} proposed a D2Dassist data transmission protocol (LSD) that helps integrity, data confidentiality and mutual authentication through applying cryptographic primitives. However, other contextual privacy requirements have not been met as advertised.

Yang et al. \cite{yang2018privacy} structured a privacy-preserving e-health system, which is a fusion of IoT, big data and cloud storage. 

Yang et al. \cite{yang2018privacy} has configured a privacy-preserving e-health system that is a fusion of big data, big data and cloud storage. They provide access control as well as content privacy based on attribute-based encryption.

The main benefits of federated learning are
\begin{itemize}
    \item FL allows devices such as smartphones to learn a shared prediction model cooperatively while keeping the training data on the device rather than uploading and storing it on a central server. It also protects privacy because devices will be unable to access or save the data shared with them.  As described in the paper by Bonawitz et al. \cite{2019arXiv190201046B}, FL can facilitate data sharing with multiple entities on the network to improve data security and availability in the network .
    \item Moves model training to the edge, incorporating devices like smartphones, tablets, and the Internet of Things and institutions like hospitals that must adhere to strict privacy rules. Keeping personal data local has a significant security benefit.
    \item Real-time prediction is possible since prediction takes happen on the device itself. FL reduces the time delay created by transmitting raw data to a central server and then delivering the results back to the device. It also improves privacy because the training is done locally rather than centrally in one location. To increase model performance, the model is then shared with numerous peers and dispersed over the network.
    \item The prediction procedure works even with no internet connection since the models are stored on the device.
    \item FL reduces the amount of hardware infrastructure required. FL models require very little technology, and what is accessible on mobile devices is more than adequate.
\end{itemize}

The road map of the chapter is organized as follows. we explain federated learning in Section \ref{sec:fed_learning}. Section \ref{sec:dataset} contains dataset and experiment. In Section \ref{sec:experiments}, we present an extensive of privacy analysis and performance evaluation. Section \ref{sec:threats_to_val} shows threats to validity. Lastly, we conclude the chapter at Section \ref{sec:conclusion}. 

\section{Federated Learning Types}\label{sec:fed_learning}
Federated Learning (FL) broadly can be categorized into three types, such as horizontal, vertical, and hybrid \cite{yang2018privacy} based on the data partition. FL is immensely helpful for building models where data is shared across different domains. The hybrid FL is based on transfer learning, while horizontal FL leverages the data parallelism and the vertical FL leverages model parallelism \cite{pan2009survey}. 

In the following sections, we will provide different federate learning and aggregation methods.

\subsection{Horizontal Federated Learning}
\begin{figure}[htbp!]
    \centering
    \includegraphics[width=1.0\linewidth]{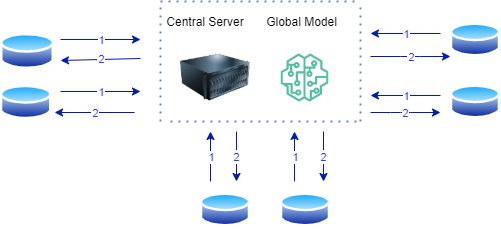}
    \caption{Horizontal FL overview}
    \label{fig:horizontal}
\end{figure}

Horizontal FL, which is mentioned to homogeneous FL \cite{haddadpour2019convergence}, is defined as the circumstances in which dataset on the devices share the same features space but are different in instances. 
Basically, horizontal FL splits the dataset horizontally, then subtract the part of the data where the user attributes are the same. Still, the users are not exactly the same for training. In this way, the user sample size can be increased by using horizontal FL. In horizontal FL, as shown in \ref{fig:horizontal}, all parties calculate and load local gradients and then the central server aggregates to provide a global model. Homomorphic encryption \cite{aono2017privacy}, secure aggregation \cite{chen2018privacy} and differential privacy \cite{mcmahan2017communication} can provide security of the transferring process of gradients in horizontal FL. Figure \ref{fig:horizontal} shows a typical Horizontal FL overview.

\subsection{Vertical Federated Learning}
Vertical FL \cite{yang2019federated} is referred as heterogeneous FL \cite{yu2020heterogeneous}, in which users’ training data share the same sample space but have different feature spaces. 

\begin{figure}[htbp!]
    \centering
    \includegraphics[width=1.0\linewidth]{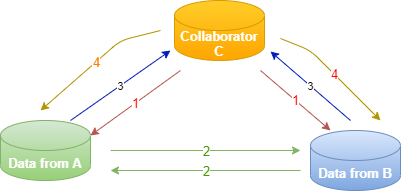}
    \caption{Vertical FL overview}
    \label{fig:vertical}
\end{figure}

Vertical FL involves dividing datasets vertically and removing the part of the data for training where the users are the same but the user characteristics are different. As a result, vertical FL has the potential to enhance the feature dimension of training data. In practice, several common solutions for vertical division of data concerns have been implemented in the literature, such as safe linear regression, classification, and gradient descent \cite{wan2007privacy}. Figure \ref{fig:vertical} shows a typical Vertical FL overview.

In the Figure \ref{fig:vertical}, arrow 1 refers sending public key, 2 is exchanging encrypted value, 3 is sending encrypted result and 4 is updating the model.

\subsection{Aggregation Algorithms}

Different aggregation techniques that compound the local model updates from all the clients participating in the training cycle have been presented in this area \cite{nilsson2018performance}. Aggregation algorithms may be divided into three types: centralized, hierarchical, and decentralized \cite{lian2017can}.

\subsubsection{Centralized Aggregation}
In the centralized aggregation configuration, a single parameter server provides a global model that calculates the average gradients gathered from clients and then coordinates their updates using a centralized aggregation algorithm. Federated averaging changes in each implementation of FL in terms of the pre-config parameters. We have presented primarily used centralized aggregation algorithms. 

Federated Averaging (FedAvg) is a primitive and most extensively applied Federated Learning algorithm introduced in Google’s implementation of FL, based on stochastic gradient descent (SGD) optimization algorithm. FedAvg includes a central server acting as a coordinator that stores the overall predictive model transmitted to a randomly selected subset of devices. Devices in the federation train the model using its local data and are responsible for sending parameter updates back to the server. Lastly, the server aggregates these updates and creates a new global model \cite{mcmahan2017communication}. 

The other algorithm is the Federated Stochastic Block Coordinate Descent(FedBCD) algorithm that has a similar principle with FedAvg, targets total rounds of communications, skipping updates for each iteration to achieve desired accuracy rate \cite{liu2019communication}. 

FedProx is a modified version of the FedAvg algorithm presented in \cite{li2018federated} to overcome heterogeneity in FL and allows multiple iterations on each compute resource while minimizing a cost function based on a local loss function and a global model. 

FedMA constructs a shared model for CNNs and LSTM based ML model updates in FL environments \cite{wang2020federated}. FedMA averages models on the central server by mapping and averaging hidden elements in neural networks, such as neurons and channels, on a layer-by-layer basis. Stochastic Controlled Averaging for FL (Scaffold) [64] reduces communication rounds by using stateful variables in distributed computing resources \cite{karimireddy2020scaffold}.

Attentive Federated Aggregation (FedAttOpt) adds an enhanced mechanism for modelling aggregation on the central server of the computed FL \cite{jiang2020decentralized}.

\subsection{Hierarchical Aggregation}
A hierarchical architecture is utilized by using multiple parameter servers that named global parameter server (GPS) and multiple region parameter servers (RPS) to decrease models' transfer time between a parameter server and computing resources \cite{yuan2020hierarchical}. Each RPS is applied in a cell base station to which computing resources can be connected with low latency. For the Hierarchical approach, various algorithms are proposed. Hierarchical Federated Learning (HFL) \cite{yuan2020hierarchical}, HierFAVG \cite{liu2020client}, LanFL \cite{yuan2020hierarchical} and HFEL \cite{luo2020hfel} are applied to perform model aggregation. 

\subsection{Decentralized Aggregation}
Dependency on the central server is excluded to enable the model collection, and peer-2-peer topology is followed for communication in a decentralized approach \cite{vanhaesebrouck2016decentralized}. Any global model is not seen, and each computing resource develops its model to share information with the other neighbours. For decentralized aggregation, various algorithms are proposed. In the decentralized SGD model (D-SGD), each computing resource keeps a global model's local copy and uses its neighbours' models to update the local copy. Average With-Communication (AWC) \cite{lalitha2019peer} and Average-Before-Communication (ABC) \cite{wang2019matcha} are two common types of execution of D-SGD.

\section{Dataset}\label{sec:dataset}
\begin{figure}[htbp!]
    \centering
    \includegraphics[width=0.9\linewidth]{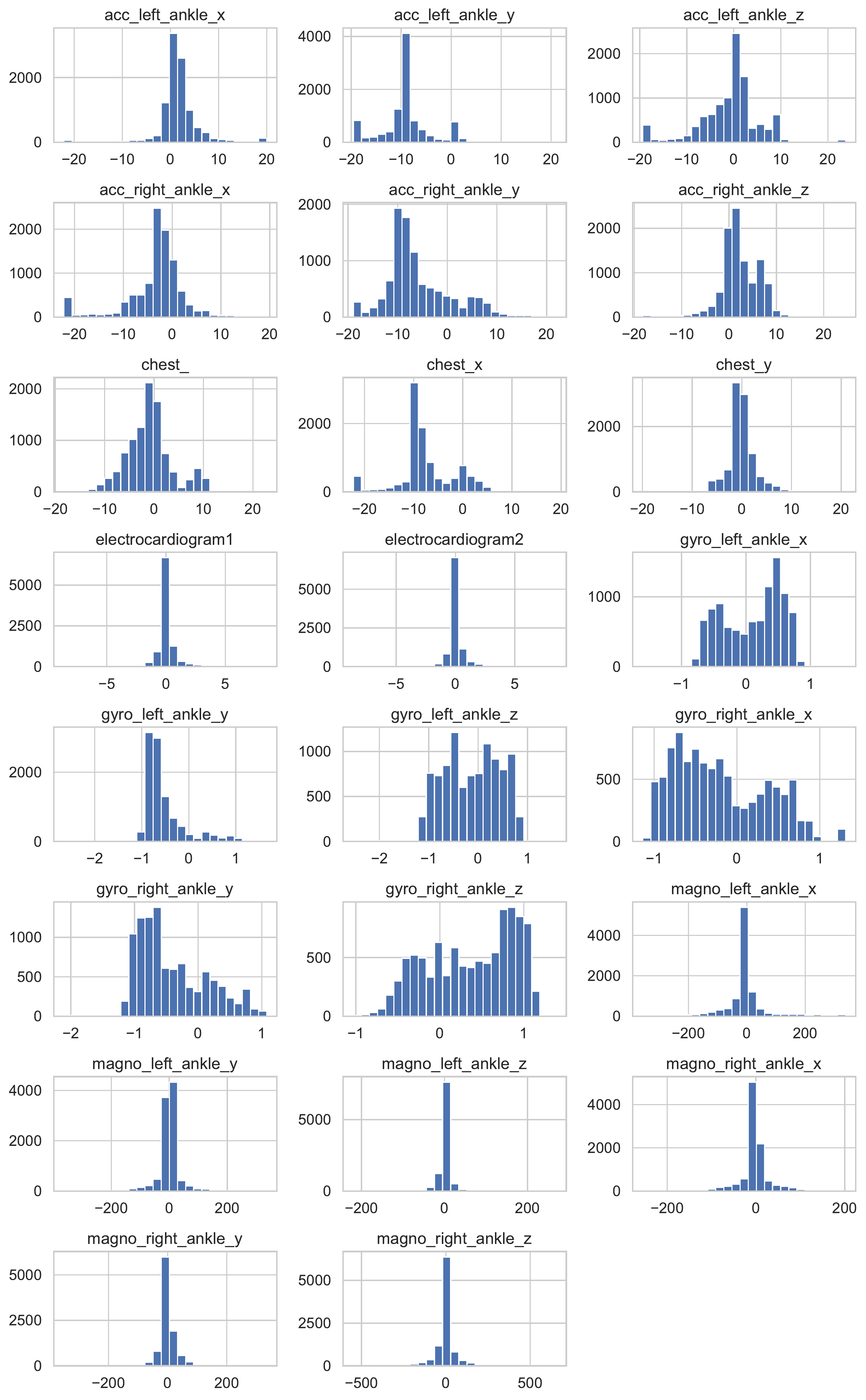}
    \caption{Columns histogram values}
    \label{fig:col_hist}
\end{figure}

In this work, we used an publicly available Mobile Health (MHEALTH) dataset \cite{10.1007/978-3-319-13105-4_14} of UCI Machine Learning Repository \footnote{http://archive.ics.uci.edu/ml/datasets/mhealth+dataset}.
The MHEALTH (Mobile HEALTH) dataset contains recordings of body motion, and vital signs were taken while 10 volunteers of various profiles engaged in multiple physical activities. Sensors are installed on the right wrist, subject's chest, and left ankle to track the movements of different body components, including acceleration, rate of rotation, and magnetic field orientation. The sensor on the chest can also take 2-lead electrocardiogram (ECG) readings, which may be utilized for basic cardiac monitoring, testing for different arrhythmias, or examining the effects of exercise on the ECG. Figure \ref{fig:col_hist} shows the column values' histogram distribution. According to the figure, almost all columns have a normal distribution.

\section{Evaluation} \label{sec:experiments}
Section \ref{sec:experiments} covers the findings about experiments about DI model for IoT devices. In Section \ref{sec:exp_design}, we present the experiment design, followed by experiment execution in Section \ref{subsec:execu}, and results in Section~\ref{subsec:results}. 

\subsection{System Overview}
\begin{figure}[htbp!]
    \centering
    \includegraphics[width=1.0\linewidth]{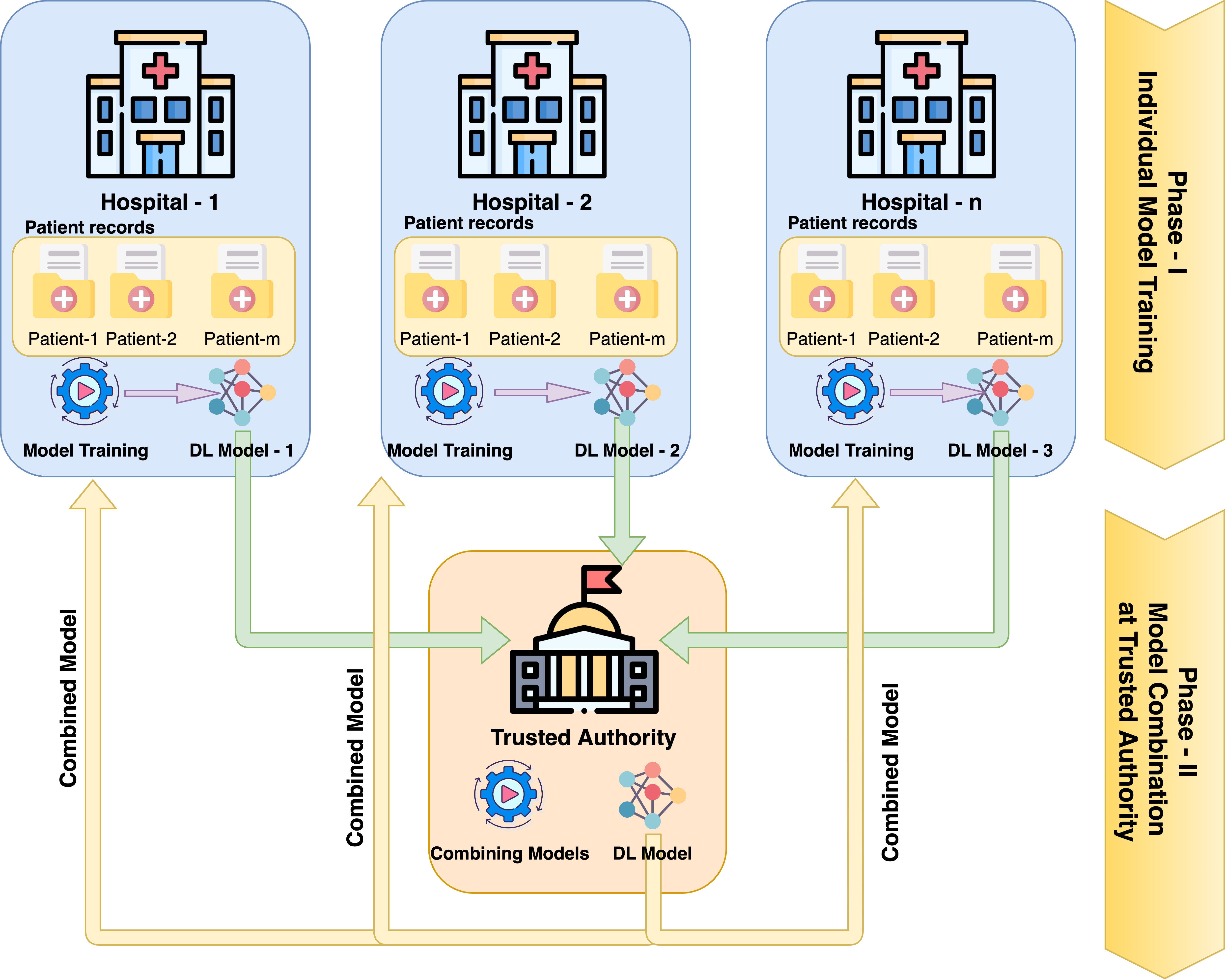}
    \caption{System overview}
    \label{fig:system-overview}
\end{figure}

The main aim of this work is to build a DL model for different hospitals based on their corresponding patient records. The proposed federated learning-based approach mainly consists of two phases: (Phase I) the model building at individual hospital phase, and (Phase II) the model combination at a trusted authority (Figure \ref{fig:system-overview}).

As shown in Figure \ref{fig:system-overview}, Phase I is responsible for data pre-processing and model building with input dataset, $\mathcal{D}$ in each hospital. A standard DL training is performed at this phase, and categorical cross-entropy loss functions are used in weight optimization with the exactly same DL architecture (i.e. number of neurons at each layer, number of layers, optimization method and activation functions). The primary outcome in each hospital is their own model, which is the model with the best prediction accuracy and lowest loss over the entire patient records.

In Phase II, the trusted authority collects all models from all hospitals. Figure \ref{fig:system-overview} illustrates multiple models with $n$ different hospitals (\textit{Hospital-1}, \textit{Hospital-2}, ... \textit{Hospital-n}), which are generated from their private datasets. The trusted authority is responsible for combining all layers and neurons (i.e. mean value of each weight). Then the trusted authority sends the combined model to each hospital for the retraining process. This process continues until there is no change in model weights. After this stage, the combined model is given to all hospitals and used as a prediction model.

Algorithm \ref{alg:fedavg} shows the pseudocode implementation of the overall process.

\begin{algorithm}[htbp!]
\DontPrintSemicolon
  \KwIn{  $T$ is the number of clients in FL, $\mathcal{D}_t = \{(X_t, \mathbf{y}_t)| X_t \in \mathbb{R}^{m \times n}, \mathbf{y}_t \in \mathbb{R}^m \}$ is the local dataset at client $t$; $L$ is the number of layers in the model $h$, $k$ is the number of iteration.}
  \KwOut{The final privacy-preserving model $h$ }
  \tcp{Randomly initialize the final model's weights}
   $h \gets rand(\mathbf{w})$ \;
   \tcp{Send the randomly initialized model to each client}
   \ForEach{$t \in T$}{
      $send\_model\_to\_client(h,t)$\;
   }
   \tcp{Iterative training}
   \For{$i=0;i<k;i++$}{
       \tcp{\textbf{\large Models training stage at each client}}
       \tcp{At each client $t$, perform a model building, $h_t$, using its local dataset, $\mathcal{D}_t$}
       \ForEach{$t \in T$}{
          \tcp{Train the local model}
          $h_t \gets train(X_t, \mathbf{y}_t) $ \;
          \tcp{Send the trained model to the central authority}
          $send\_model(h_t)$ \;
       }
       \tcp{\textbf{\large Models merging stage}}
       \tcp{Iterate each layer in the models}
       \ForEach{$l \in L$}{
          \tcp{Calculate the mean value of the layer $t$ weight vector of each model.}
          $h^{(t)} \gets mean(h_t^{(l)})$\;
        }
        \tcp{Send the merged model to each client}
        \ForEach{$t \in T$}{
          $send\_model\_to\_client(h,t)$\;
        }
    }
return $h$ 
\caption{Centralized merging of privacy-preserving model. }
\label{alg:fedavg}
\end{algorithm}

\subsection{Experiment Design}\label{sec:exp_design}
\begin{figure}[!htb]
    \centering
    \includegraphics[width=1.0\linewidth]{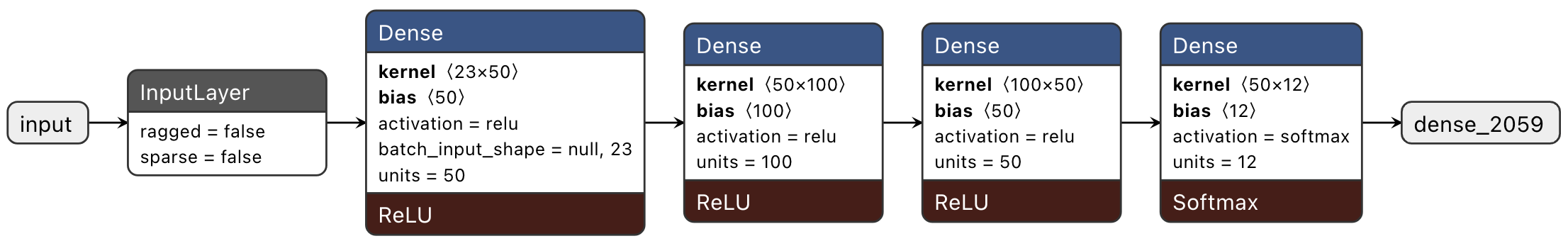}
     \caption{Dl model architecture in each IoT devices.}
	\label{fig:keras-model-overview}
\end{figure}

We designed a series of experiments by uniformly varying 5 different number of devices, $t \in [3,5,10,15,30]$. Each experiment was executed 30 times, results of which, corresponding to each $t$, are averaged to smooth the plotting. We selected the best hyper-parameters for the DL models for the dataset we used in the experiments using a simple grid search. It turned out that the best hyper-parameters are the Rmsprop optimization with a learning rate of 0.01. Figure \ref{fig:keras-model-overview} shows DL models for each datasets.

\subsubsection{Research Questions}\label{sec:rqs}

We aim to explore privacy-preserving of medical IoT devices, explore federated learning to protect privacy. We formed the following Research Questions (RQs) and designed the experiments to answer them: 
\begin{itemize}
    \item \textbf{RQ1}: Is federated learning based privacy protection applicable for medical IoT devices? Studying this RQ is necessary as a positive answer to it motivates us to further investigate whether federated learning can help to protect the privacy.
    \item \textbf{RQ2}: Is there any relationship between prediction performance and number of devices while protecting the privacy of medical IoT devices. RQ2 helps us assess the quality of a DL model's prediction while protecting the privacy of the dataset. 
\end{itemize}

\subsection{Experiment Execution} \label{subsec:execu}

All the experiments were performed using Python scripts and ML libraries: Keras, Tensorflow, and Scikit-learn, on the following machine: 2.8 GHz Quad-Core Intel Core i7 with 16GB of RAM. 

The dataset has been divided into two parts: 20\% test dataset and 80\% training dataset. In the training dataset, we randomly distributed subsets of the dataset to the simulated IoT devices and used them to build the DL model. The number of epochs was set to 50, and the epoch size was fixed for all the models in each IoT device. 

\subsection{Prediction Performance Metrics}
In this study, we used four metrics (overall prediction accuracy, recall, precision, and $F_1$ score) that are common measurement metrics to evaluate classification accuracy and to find an optimal classifier in machine learning. \cite{10.1145/1148170.1148176,manning2008introduction,Makhoul99performancemeasures}.

Precision is defined as the fraction of retrieved samples that are relevant:
\begin{equation}
    Precision = \frac{Correct}{Correct + False}
\end{equation}

Recall is defined as the fraction of relevant samples that are retrieved:
\begin{equation}
    Recall = \frac{Correct}{Correct + Missed}
\end{equation}

The $F_1$ score is defined as the harmonic mean of precision and recall:
\begin{equation}
    F_1 = 2 \times \frac{Precision \times Recall}{Precision + Recall}
\end{equation}

\subsection{Results}\label{subsec:results}
\subsubsection{Results for RQ1}\label{sec:res_rq1}
\begin{figure}
    \centering
    \begin{subfigure}[b]{0.49\linewidth}
         \centering
         \includegraphics[width=1\linewidth]{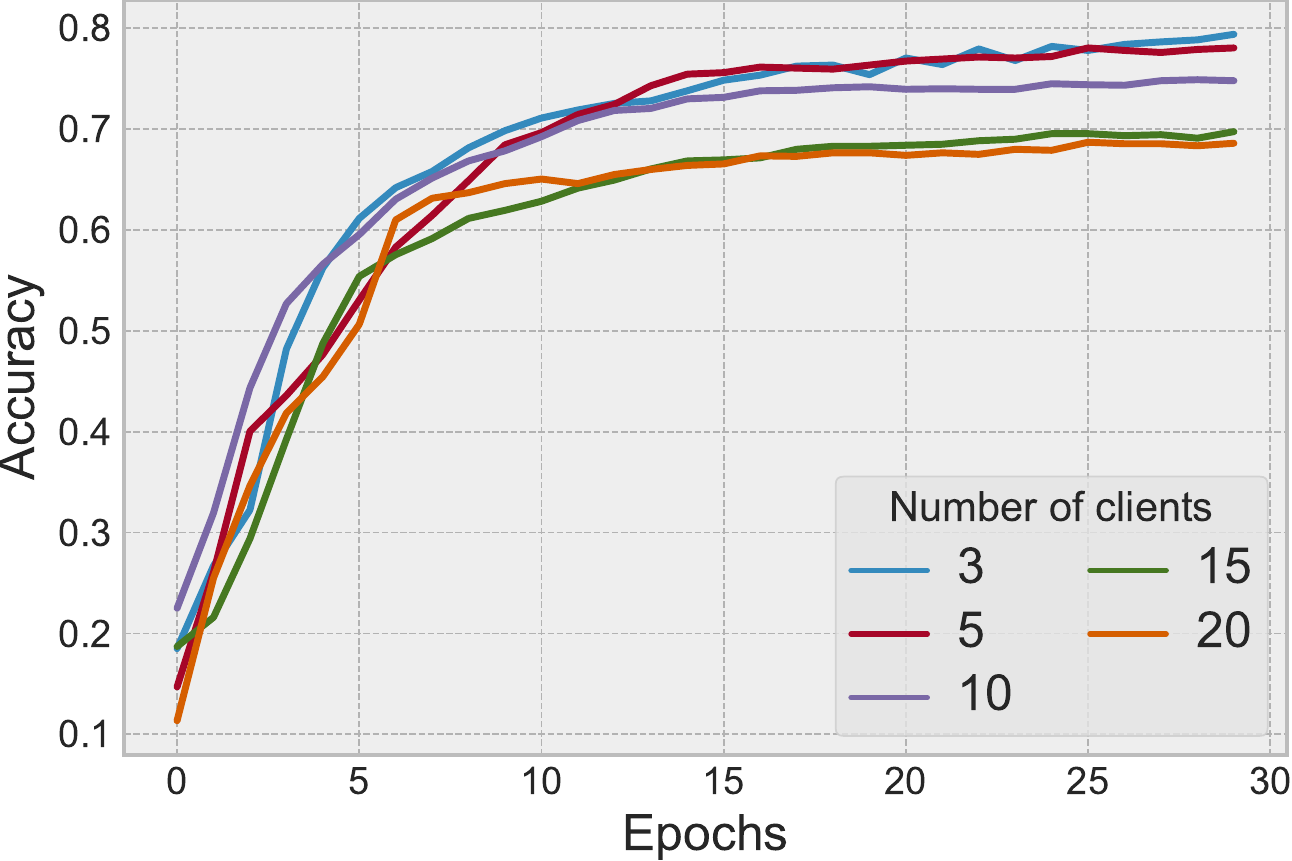}
	\caption{Video conferencing system}
	\label{fig:Accuracy}
     \end{subfigure} \hfill
     \begin{subfigure}[b]{0.49\linewidth}
         \centering
         \includegraphics[width=1\linewidth]{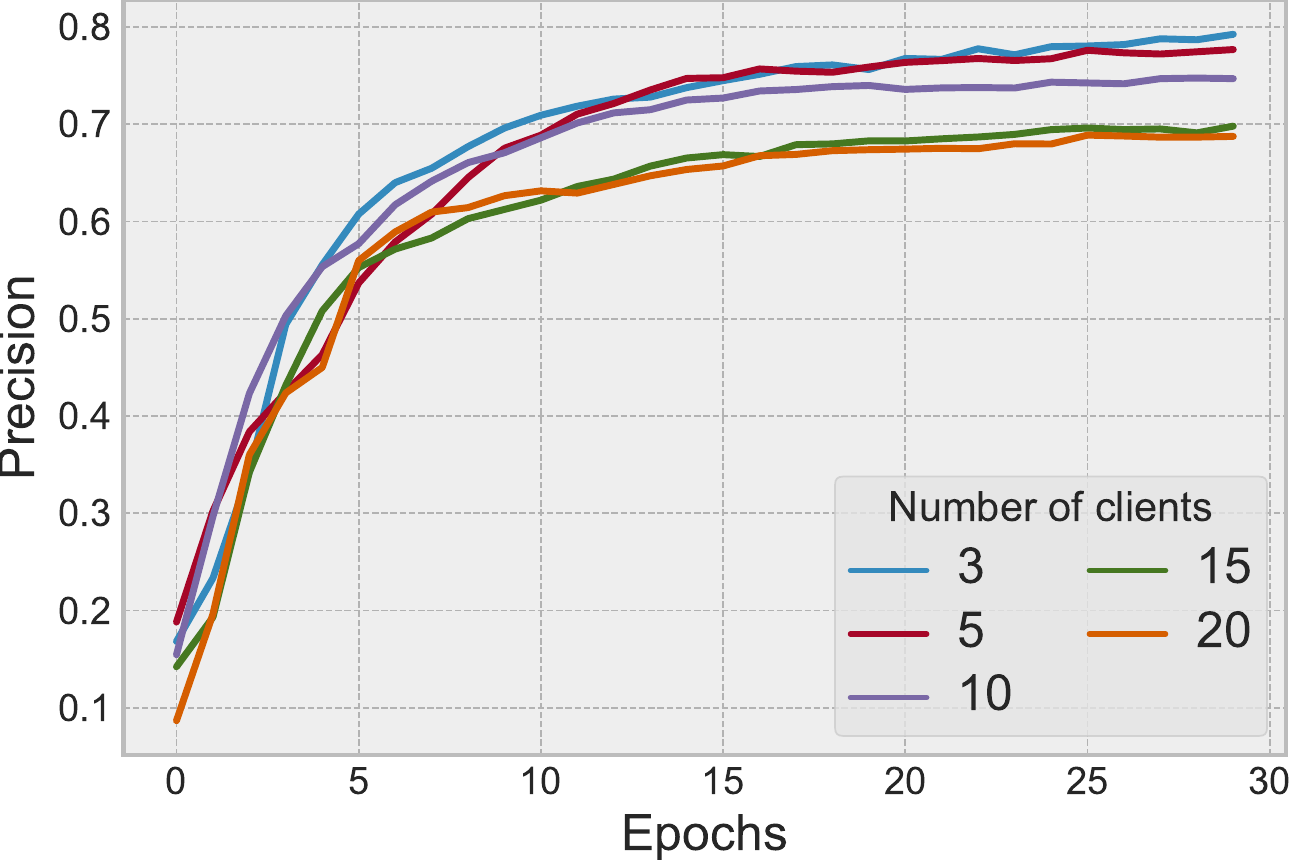}
	\caption{Video conferencing system}
	\label{fig:Precision}
     \end{subfigure} \hfill
     \begin{subfigure}[b]{0.49\linewidth}
         \centering
         \includegraphics[width=1\linewidth]{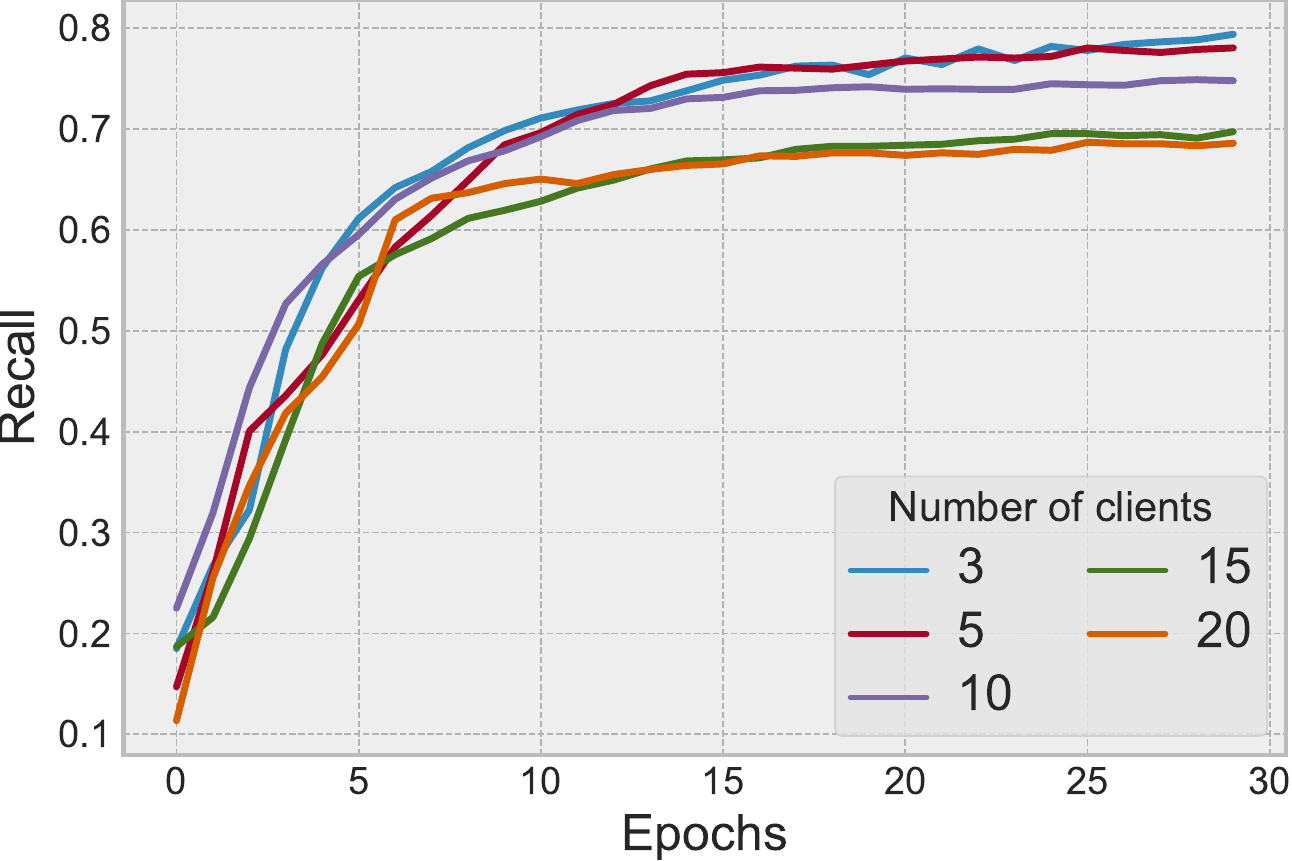}
	\caption{Video conferencing system}
	\label{fig:Recall}
     \end{subfigure} \hfill
     \begin{subfigure}[b]{0.49\linewidth}
         \centering
         \includegraphics[width=1\linewidth]{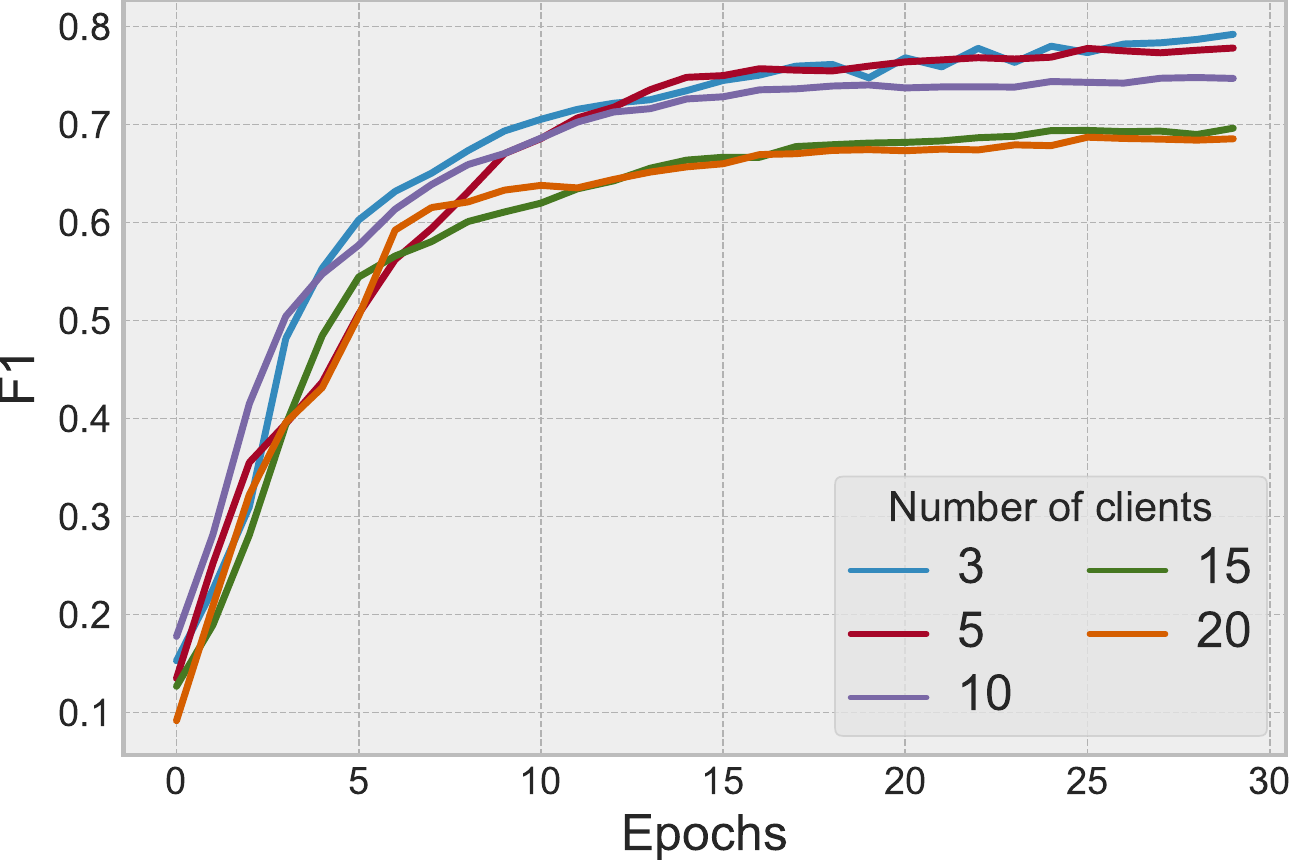}
	\caption{Video conferencing system}
	\label{fig:F1}
     \end{subfigure} \hfill
    \caption{The DL model's prediction performance with different number of IoT clients.}
    \label{fig:accuracy}
\end{figure}

To answer this RQ, first, we obtain values of the prediction performance metrics (i.e., accuracy, precision, recall, and $F_1$) of the DL models, which are shown in Figure \ref{fig:system-overview}. Figure \ref{fig:accuracy} delivers the prediction performance for the dataset in terms of accuracy, precision, recall, and $F_1$. As shown in the figure, there is a negative relationship between the number of clients in the federated learning and prediction accuracy. With the number of clients, a decrease is observed in the final model classification performance. Thus, as shown in the figure, the DL model performance drops considerably as the number of clients is increased. Moreover, although the total number of clients varies and the final model classification accuracy is almost equal to the base model's accuracy. The negative relationship in the model prediction performance is related to the fact that the number of clients can also reduce the dataset size in each hospital. 

\subsubsection{Results for RQ2}\label{sec:res_rq2}
\begin{figure}[htbp!]
    \centering
    \includegraphics[width=1.0\linewidth]{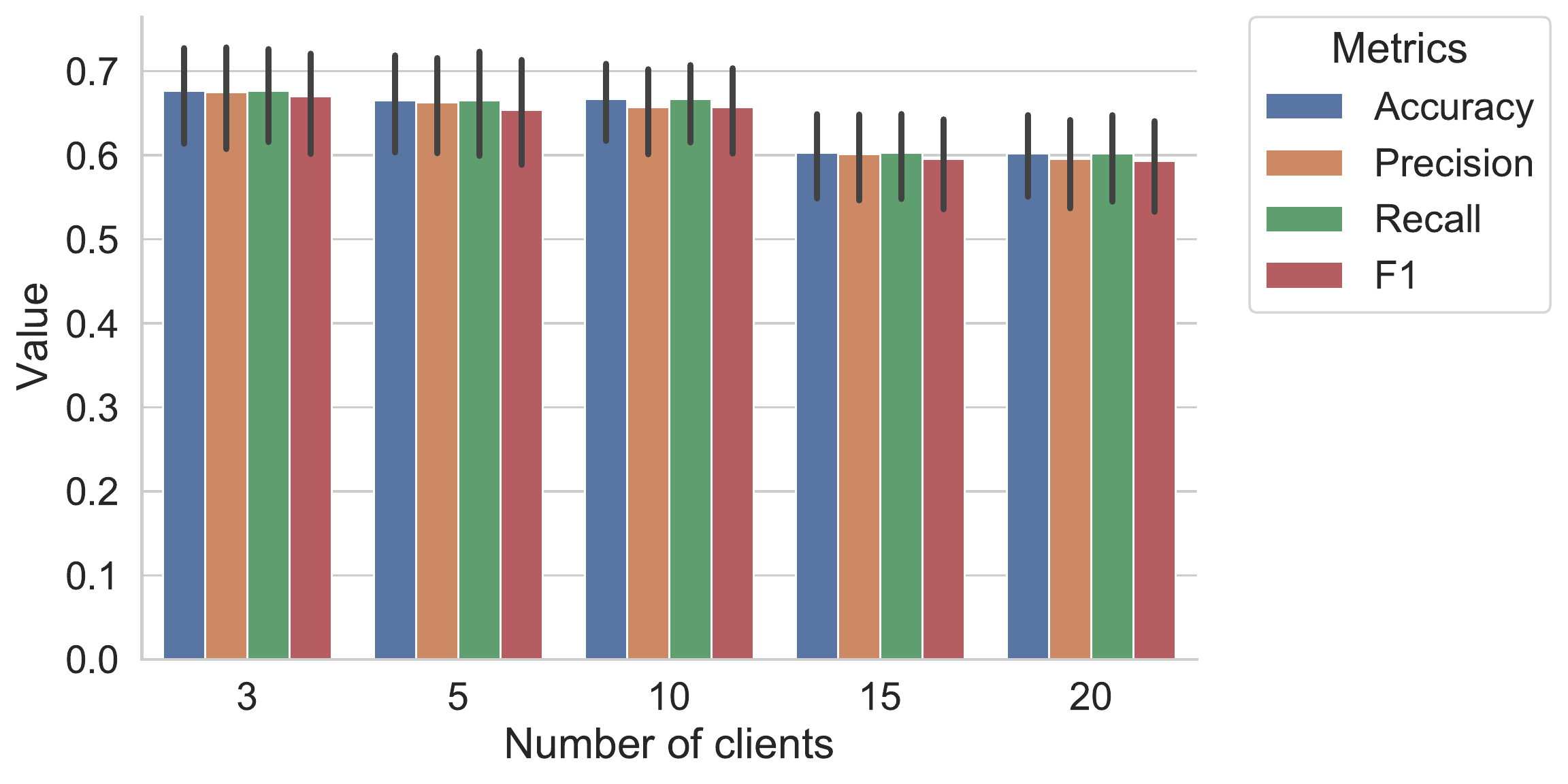}
    \caption{Prediction performance with the different number of clients}
    \label{fig:pred_perf_clients}
\end{figure}

Figure \ref{fig:pred_perf_clients} shows the accuracy, precision, recall and $F_1$ metrics with a different number of clients in the federated learning environment. As illustrated in the figure, the number of clients in the federated learning environment and prediction accuracy has a negative relationship. The final model's classification performance appears to be decreasing as the number of clients grows. The figure shows that with increasing the number of clients, the DL model is becoming less accurate. However, with increasing the number of clients, the number of patients in the dataset is also decreasing. 

\section{Threats to Validity}\label{sec:threats_to_val}

A key \textit{external validity} threat is related to the generalization of results. In our experiments, we used only one medical IoT dataset. We also definitely require more case studies to generalize the results. Besides, the dataset reflects ECG readings from human body.

Our key \textit{construct validity} threat is related to selecting the number of IoT devices that do not precisely show the DL model's prediction performance. Nevertheless, note that the number of clients varies, and the DL models' prediction performance is very close to each others. In the future studies, we will conduct more empirical studies to systematically investigate the effects of the number of clients in federated learning to the DL models' prediction performance.

\section{Conclusion}\label{sec:conclusion}
Federated learning was mentioned as a new way to provide privacy preservation. The primary purpose of this study is to create a DL model for different hospitals based on the relevant patient records. Federated learning-based privacy protection was proposed for medical IoT devices. This study contributes to explore privacy-preserving of medical IoT devices and federated learning to protect privacy. We point out that the accuracy of the Bayesian NN model is roughly equal to the accuracy of the prediction performance of the base model. We indicate that the number of clients varies and the prediction performance of the DL models is very close to each other. 
 
 As future research, we will implement partial and somewhat homomorphic encryption-based schemes to protect personal and sensitive data in IoT applications.

\bibliographystyle{ieeetr}
\bibliography{references}
\end{document}